\documentclass[pra,a4paper,preprint,showpacs,superscriptaddress]{revtex4-2}
\usepackage{amsmath,amscd,amsfonts,amssymb, amsbsy, color,bm}
\usepackage{graphicx}
\begin{document}
	\title{Margenau-Hill operator valued measures and joint measurability}
	\author{Seeta Vasudevrao}
	\affiliation{Department of Physics, Bangalore University, Bangalore-560056, India}
	\author{H. S. Karthik}
	\affiliation{International Centre for Theory of Quantum Technologies, University of Gdansk, Gdansk, Poland}
	\author{I. Reena}
	\affiliation{Department of Physics, Bangalore University, Bangalore-560056, India}
	\author{Sudha} 
	\affiliation{Department of Physics, Kuvempu University, 
		Shankaraghatta-577 451, Karnataka, India}
	\email{tthdrs@gmail.com}
	\author{A. R. Usha Devi}
	\affiliation{Department of Physics, Bangalore University, Bangalore-560056, India}
	\affiliation{Inspire Institute Inc., Alexandria, Virginia, 22303, USA. }
\begin{abstract}
We employ the Margenau-Hill (MH) correspondence rule for associating classical functions with quantum operators to construct  quasi-probability mass functions. Using this we obtain the  fuzzy one parameter {\em quasi measurement operator } (QMO) characterizing the incompatibility of  non-commuting spin observables of qubits, qutrits and 2-qubit systems. Positivity of the fuzzy MH-QMO places upper bounds on the associated unsharpness parameter.  This serves as a sufficient condition for measurement incompatibility of spin observables. We assess the amount of {\em unsharpness}  required for joint measurability (compatibility) of the  non-commuting qubit, qutrit and 2-qubit observables. We show that the {\em degree of compatibility} of a pair of orthogonal qubit observables agrees perfectly with the necessary and sufficient conditions for joint measurability. Furthermore, we obtain analytical upper bounds on the unsharpness parameter specifying the range of joint measurability of  spin components of qutrits and  pairs of orthogonal spin observables of a 2-qubit system. Our results  indicate that the measurement incompatibility of spin observables increases with  Hilbert space dimension.  
	  \end{abstract}
\maketitle
\section{Introduction}	
In the classical framework, physical observables are all compatible and they can be measured jointly. In contrast, measurement of observables, which do not commute, are declared to be incompatible in the quantum scenario. The notion of compatibility of measurements is limited to that of commutativity of the observables only when one restricts to projective valued (PV) measurements. An extended notion of incompatibility emerges in the generalized framework of {\em fuzzy} measurements using positive operator valued measures (POVMs). Quantum measurement incompatibility offers a conceptual way to separate quantum and classical features~\cite{Busch,Hein,Hein2}.

In a different direction,  Wigner's quasi-phase-space distribution function approach~\cite{Wig} brings out differences between quantum and classical phase-space descriptions. In particular, the quantum phase-space distribution differs from its classical counterpart in that it  can assume negative values or it is more singular than the Dirac delta function distribution~\cite{Simon_Mukunda}. Hence, it cannot be treated as a genuine probability distribution function. For this reason, Wigner's quantum phase-space distribution function is referred to as {\em quasi-probability distribution}. 

Based on Wigner's approach, it is possible to reproduce expectation values of quantum observables as  phase-space averages of corresponding classical functions.   Besides Wigner distribution function, which makes use of Weyl's operator correspondence rule~\cite{Weyl}
$$q^k\,p^l\, \longleftrightarrow\, \frac{1}{2^k}\sum_{r=0}^{k}\, \frac{k!}{r!\, (k-r)!}\,\hat{q}^r\, \hat{p}^l\, \hat{q}^{k-r}, \ k, r=1,2,\ldots$$
to associate classical functions $f(q,p)$ of canonical position and momentum variables $q,\ p$   with quantum operators $\hat{f}(\hat{q},\, \hat{p})$, various other distribution functions have also been developed by designing suitable operator correspondence rules
~\cite{Review,AKR_Pramana,MH,others,gr,puri1,puri2}. The phase-space distribution formalism, initiated by Wigner for canonical position and momentum observables, has also been extended to develop discrete probability mass functions for spin observables~\cite{MH,others,gr,puri1,puri2}.  

In this work we employ the Margenau-Hill (MH) operator correspondence rule~\cite{Review,AKR_Pramana,MH,others,gr,puri1,puri2}
$$x_1^k\,x_2^r\, \longleftrightarrow\, \frac{\hat{X}^k_1\, \hat{X}^r_2+\hat{X}^r_2\, \hat{X}^k_1 }{2},\ \ k, r=1,2,\ldots$$
for associating classical functions $f(x_1,x_2)$ with quantum operator functions $\hat{f}(\hat{X}_1,\hat{X}_2)$ of non-commuting observables $\hat{X}_1$, $\hat{X}_2$ in a $d$-dimensional Hilbert space and  construct  quasi-probability mass functions $P_{{\rm MH}}(x_1,\,x_2)$ such that 
$$\sum_{x_1,x_2=1}^{d}\,P_{{\rm MH}}(x_1,\,x_2) f(x_1,x_2)\equiv {\rm Tr}\left[\hat{\rho}\, \hat{f}(\hat{X}_1,\hat{X}_2)\right]=\left\langle  \hat{f}(\hat{X}_1,\hat{X}_2) \right\rangle$$ where $\hat{\rho}$ denotes the density operator. Based on this approach we  explicitly construct the fuzzy one parameter MH {\em quasi measurement operator} (QMO) $\{\hat{G}_Q(x_1,\, x_2;\,\eta),\ 0\leq \eta\leq 1\}$ corresponding to the quasi mass functions $P_{{\rm MH}}(x_1,\,x_2)$, where the parameter $\eta$ denotes {\em fuzziness} of measurement. The observables  $\hat{X}_1$, $\hat{X}_2$ are {\em compatible or jointly measurable} if the MH-QMO turns out to be a legitimate parent POVM~\cite{Busch,Hein,Hein2}.  In Sec.~2 we give a brief outline on {\em measuremet incompatibility} of a set of finite outcome quantum observables. The terms {\em compatibility} and {\em joint measurability} are used interchangeably throughout our discussion.

\section{Joint measurability of non-commuting observables} 
We start with a brief description of generalized measurements and the notion of joint measurability: 
\begin{itemize}
\item[$\bullet$] A POVM  is a set $\left\{\hat{E}(x)\right\}$ of positive semidefinite operators acting on a Hilbert space of dimension $d$, satisfying the normalization condition 
$$\sum_{x=1}^d\,\hat{E}(x) = \hat{I}_d,$$ 
 where $\hat{I}_d$ denotes the identity operator.  The element $\hat{E}(x)$ of the POVM is associated with the measurement outcome $x=1,2,\ldots , d$. 

\item[$\bullet$] Incompatibility  of measurements of a set of POVMs associated with the measurement of non-commuting observables is defined as the {\em impossibility of a single parent measurement} POVM from which the results of individual measurements can be registered~\cite{Busch,Hein,Hein2}. 

Let $\hat{X}_1,\,\hat{X}_2,\ldots\, ,\hat{X}_n$ denote a set of non-commuting observables on a $d$-dimensional Hilbert space with finite measurement outcomes denoted by $x_1,\,x_2,\ldots,\, x_n$ respectively. Let  $E_k\equiv \{\hat{E}_\eta(x_k) \}, \  k=1,2,\ldots n$ denotes the POVM associated with the generalized  measurements of the observable $\hat{X}_k$, with the real parameter $0\leq \eta\leq 1$ characterizing {\em unsharpness} or {\em fuzziness}~\cite{LSW}. 

In principle, the POVMs  $\hat{E}_1,\,\hat{E}_2,\ldots\, ,\hat{E}_n$  are {\em jointly measurable}  if there exists a {\em parent}  POVM  $\{\hat{G}(x_1,x_2,\ldots , x_n;\,\eta); 0\leq \eta \leq 1\}$ acting on the  $d$-dimensional Hilbert space, which obeys the following properties~\cite{Busch,Hein,Hein2,nc}:
\begin{itemize} 
	\item[(i)] \hskip 0.25in   $0\leq\hat{G}(x_1,x_2,\ldots , x_n; \eta)\leq \hat{I}_d$.    
	\item[(ii)]   $\displaystyle\sum_{x_1,x_2,\ldots\, x_n}\, \hat{G}(x_1,x_2,\ldots , x_n; \eta)= \hat{I}_d$.
	\item [(iii)] \hskip 0.25in $\hat{E}_\eta(x_k)=\displaystyle\sum_{x_1,x_2,\ldots, x_{k-1},\, x_{k+1},\ldots, x_n}\, \hat{G}(x_1,x_2,\ldots , x_n; \eta)$. 
\end{itemize} 
\end{itemize}
Several  measures to quantify incompatibility of measurements in terms of a single real parameter $\eta\in[0,1]$ have been proposed in the literature~\cite{Busch,Hein,Hein2,njp19}. 
\begin{itemize}
\item[$\bullet$] Heinosaari et. al~\cite{Hein2} consider {\em degree of compatibility} of the POVMs $\hat{E}_1,\,\hat{E}_2,\ldots\, ,\hat{E}_n$  as the {\em supremum value} (least upper bound)  of a single real parameter $\eta\in[0,1]$ which ensures that the fuzzy POVMs  $\hat{E}_k=\{\hat{E}_{\eta}(x_k)\}, k=1,2,\ldots n$   are all jointly measurable.  
\item[$\bullet$] A real parameter $\eta\in[0,1]$ is used as an index  to quantify  {\em minimum} amount of {\em noise}  added so that the set of  observables are compatible (see Ref.~\cite{njp19} and references therein).  The supremum  $\eta^*$ of this real parameter  is called {\em the incompatibility robustness}. This interpretation  comes from the convex-geometric structure of POVMs~\cite{Hein2,njp19}.
\item[$\bullet$]  The region $0\leq\eta\leq \eta^*$ in which a bonafide parent POVM  $\{\hat{G}(x_1,x_2,\ldots , x_n; \eta)\}$ is guaranteed has been  termed as {\em the  compatibility region}~\cite{Hein2} of the POVMs $\{\hat{E}_\eta(x_k)\}, k=1,2,\ldots n$. Larger the compatibility region more compatible are the  observables. It has been realized that the region of compatibility decreases with the increase of Hilbert space dimension~\cite{Hein2,njp19}. 
\end{itemize}
For a detailed overview  on  robustness based measures describing {\em how incompatible a set of quantum observables are to specific noise models} see Ref.~cite{njp19} Also see Ref.~\cite{Hein2} for a  review on measurement incompatibility in more general {\em operational theory} or {\em  probability theory} framework.

The purpose of this work is to  construct quasi-measurement operators associated with non-commuting observables by employing the Margenau-Hill correspondence rule~\cite{Review,AKR_Pramana,MH,others,gr,puri1,puri2} and explore the compatibility region of quantum spin observables of qubits, qutrits and 2-qubit systems~\cite{ft1}.
\section{Margenau-Hill characteristic function for non-commuting observables }

Margenau-Hill (MH) characteristic function for non-commuting observables $\hat{X}_1,\, \hat{X}_2,\ldots, \hat{X}_n$ is defined by~\cite{MH,others}
\begin{eqnarray}
	\label{mhch}
	\phi_{\rm MH}(u_1,u_2,\ldots, u_n)&=& \left\langle\, \left\{\,  e^{i\, \hat{X}_1\, u_1}\,e^{i\, \hat{X}_2\, u_2}\ldots\, e^{i\, \hat{X}_n\, u_n}  \right\}_{\rm Sym}\, \right\rangle \nonumber \\ 
	&=& 	\frac{1}{n!}\,\left\langle\, \sum_{P}\, \left\{\,  e^{i\, \hat{X}_1\, u_1}\,e^{i\, \hat{X}_2\, u_2}\ldots\, e^{i\, \hat{X}_n\, u_n}\, \right\}\, \right\rangle \nonumber \\ 
	&=& \frac{1}{n!}\, \sum_{P}{\rm Tr}\left(\hat{\rho} \, \left\{\,  e^{i\, \hat{X}_1\, u_1}\,e^{i\, \hat{X}_2\, u_2}\ldots\, e^{i\, \hat{X}_n\, u_n}\, \right\}\right) \nonumber \\ 
	&=& \sum_{x_1,x_2,\ldots , x_n}\, P_{\rm MH}(x_1,x_2,\ldots , x_n)\,\,  e^{i\, \sum_{k}\, x_k\, u_k}. 
\end{eqnarray}
where $\{\,\cdot\, \}_{\rm Sym}$ denotes symmetrization and  $\sum_{P}$ stands for summation over $n!$ permutations of 
the factors $e^{i\, \hat{X}_1\, u_1},\,e^{i\, \hat{X}_2\, u_2},\ldots\, ,e^{i\, \hat{X}_n\, u_n}$. 
\begin{enumerate}
	\item The set of real numbers $\{x_1,x_2,\ldots , x_n\}$ denote eigenvalues i.e., measurement outcomes of the observables $\hat{X}_1,\, \hat{X}_2,\ldots, \hat{X}_n$.
	\item The map  $\hat{\rho}\mapsto P_{\rm MH}(x_1,\,x_2,\ldots,\,x_n)$ assigns MH quasi-probability mass function  $P_{\rm MH}(x_1,x_2,\ldots , x_n)$ with any arbitrary quantum state $\hat{\rho}$ and  is obtained~\cite{MH,gr} by taking  discrete Fourier transform of the characteristic function $\phi_{\rm MH}(u_1,u_2,\ldots, u_n)$ .  
	\item The MH probability mass function is a real, normalized function satisfying the relations $P_{\rm MH}(x_1,x_2,\ldots , x_n)=  P^*_{\rm MH}(x_1,x_2,\ldots , x_n)$ and  
	$\displaystyle\sum_{x_1,\,x_2,\,\ldots,\,x_n}\, P_{\rm MH}(x_1,x_2,\ldots, x_n)~=~1$. 
	\item The marginal probability mass functions say, $P_{\rm MH}(x_k)$, is obtained by summing 
	$P_{\rm MH}(x_1,x_2,\ldots , x_{k-1}, x_k,x_{k+1},\ldots , x_n)$ over all the outcomes  except $x_k$ i.e., 
	\[ P_{\rm MH}(x_k)=\sum_{x_1,x_2\ldots,x_{k-1},\,x_{k+1}, x_n}\, P_{\rm MH}(x_1,x_2,\ldots , x_n).
	\]	
	\item The operator correspondence rule 
	\begin{eqnarray}
		\label{crule}
		x_1^{r_1}\, x_2^{r_2}\, \ldots x_{n}^{r_n} &\longrightarrow&  \{\hat{X}_1^{r_1}\, \hat{X}_2^{r_2}\, \ldots \hat{X}_{n}^{r_n}\}_{\rm Sym},  \nonumber \\
		& =&\frac{1}{n!} \sum_P\, \{\hat{X}_1^{r_1}\, \hat{X}_2^{r_2}\, \ldots \hat{X}_{n}^{r_n}\},\ \ \ r_1,\,r_2,\ldots r_n
		=0,1,2\ldots
	\end{eqnarray}
	leads to the evaluation of expectation values of  any quantum observables $\hat{f}(\hat{X}_1,\hat{X}_2,\ldots,\hat{X}_n)$ in terms of averages of corresponding classical functions 
	$ f(x_1,x_2,\ldots\, ,x_n)$ i.e.,
	\begin{equation} 
		{\rm Tr}\left[\hat{\rho}\,\hat{f}(\hat{X}_1,\hat{X}_2,\ldots,\hat{X}_n)\right]=
		\sum_{x_1,x_2,\ldots,x_n} P_{\rm MH}(x_1,x_2,\ldots,x_n)f(x_1,x_2,\ldots,x_n). 
	\end{equation} 
	\item The MH probability mass function  $P_{\rm MH}(x_1,x_2,\ldots , x_n)$ is not necessarily positive and hence is referred to as MH {\em quasi}-probability mass function. 
\end{enumerate}
\section{Margenau-Hill quasi measurement operator (MH-QMO)}
Expressing the quasi MH probability mass function $P_{\rm MH}(x_1,x_2,\ldots , x_n)$ as       
\begin{equation}
	\label{GQ}
	P_{\rm MH}(x_1,x_2,\ldots , x_n)={\rm Tr}[\hat{\rho}\,\hat{G}_{Q}(x_1,\, x_2,\, \ldots ,\, x_n)]
\end{equation}
it is possible to identify a set of  operators $\{\hat{G}_Q(x_1,\, x_2,\, \ldots ,\, x_n)\}$ which we refer to as MH {\em quasi measurement operator} (QMO) associated with the  observables $\hat{X}_1,\hat{X}_2,\ldots,\hat{X}_n$. By construction (see (\ref{GQ})) the QMO $\{\hat{G}_Q(x_1,\, x_2,\, \ldots ,\, x_n)\}$ obey 
\begin{eqnarray}
	\label{mhqdef}
	(1) && \hskip 0.25in \hat{G}^\dag_{Q}(x_1,\, x_2,\, \ldots ,\, x_n)=\hat{G}_Q(x_1,\, x_2,\, \ldots ,\, x_n)  \nonumber \\
	(2) && \hskip 0.2in \sum_{x_1,\, x_2,\, \ldots , x_n}\,\hat{G}_Q(x_1,\, x_2,\, \ldots ,\, x_n)= \hat{I}_d,\ (\mbox{Identity operator}) \nonumber \\ 
	(3) && \hskip 0.2in \sum_{x_n}\,\hat{G}_Q(x_1,\, x_2,\, \ldots ,\, x_n)=\hat{G}_Q(x_1,x_2,
	\ldots ,x_{n-1}),
	\nonumber \\
	&& \hskip 0.6in	\vdots \hskip 0.6 in \vdots \hskip 1,2in    \vdots \hskip 0.6in\vdots\nonumber \\
	&& \hskip 0.02in	\sum_{x_3,x_4,\ldots ,\, x_n}\,\hat{G}_Q(x_1,\, x_2,\, \ldots ,\, x_n)=\hat{G}_Q(x_1,x_2) \nonumber \\ 
	&& \hskip 0.02in	\sum_{x_2,\ldots, x_n}\,\hat{G}_Q(x_1,\, x_2,\, \ldots ,\, x_n)=\hat{E}(x_1).
\end{eqnarray}
As the MH joint mass function $ P_{\rm MH}(x_1,x_2,\ldots , x_n)$ is not necessarily positive the operators $\hat{G}_Q(x_1,x_2,\ldots , x_n)$ are not positive  in general. We refer to the set of operators $\{\hat{G}_Q(x_1,x_2,\ldots , x_n)\}$  as {\emph {Margenau-Hill Quasi Measurement Operator}} (MH-QMO) associated with the observables $\hat{X}_1,\, \hat{X}_2,\ldots, \hat{X}_n$. The observables $\hat{X}_1,\, \hat{X}_2,\ldots, \hat{X}_n$ are jointly measurable if 
$\{\hat{G}_Q(x_1,\, x_2,\, \ldots ,\, x_n)\}$ happens to be a legitimate parent POVM i.e., 
\begin{eqnarray}
	\label{gqp}    
	\hat{G}_Q(x_1,x_2,\ldots\, x_n)\geq 0  
\end{eqnarray} 
in addition to the defining properties (\ref{mhqdef}) of MH-QMO. The criterion given in (\ref{gqp})  serves as a sufficient condition for joint measurability of the observables $\hat{X}_1,\, \hat{X}_2,\ldots, \hat{X}_n$.  

It is pertinent to point out here that Heinosaari et. al. had alluded to the construction of an observable  (see  Sec.3.2 of Ref.~\cite{Hein2}) 
\begin{equation}
	\label{jordan}
	\hat{J}_n(\hat{E}_1,\  \hat{E}_2,\ldots,\, \hat{E}_n)=\frac{1}{n!} \displaystyle\sum_P\, \{\hat{E}_1, \hat{E}_2\ldots\, \hat{E}_n\}
\end{equation} 
associated with a set  $\hat{E}_1,\  \hat{E}_2,\ldots,\, \hat{E}_n$ of POVMs,  by generalizing the notion of {\em the Jordan product}. (In (\ref{jordan}) `P' stands for the set of all permutations). The Jordan product (\ref{jordan}) can be employed as a  joint observable whenever $\hat{J}_n(\hat{E}_1,\  \hat{E}_2,\ldots,\, \hat{E}_n)\geq 0$. While the construction (\ref{jordan}) resembles the MH operator correspondence rule given in (\ref{crule}) (see right hand side of (\ref{crule}) with $r_1,r_2,\ldots r_n=1$),  we  emphasize that the conceptual origin of  MH operator correspondence rule (\ref{crule}) is different from that of the Jordan product.     

\section{Margenau-Hill Quasi Measurement Operators (MH-QMOs) for a pair of orthogonal spin observables} 

For a pair of non-commuting observables $\hat{X}$ and $\hat{Z}$ with discrete eigenvalues $x$, $z$, the MH-QMO $\{\hat{G}_Q(x,\,z)\}$ associated with the Margenau-Hill probability distribution mass function $P_{\rm MH}(x,\,z)$ is obtained by using (\ref{GQ}): 
\begin{equation}
	P_{\rm MH}\,(x,\,z)=\mbox{Tr}\,\left[\hat{\rho}\,\hat{G}_Q\,(x,\,z)\right]=\left\langle \hat{G}_Q\,(x,\,z) \right\rangle.
\end{equation} 
It may be noted that the relations (see (\ref{mhch}), (\ref{GQ})) 
\begin{eqnarray*}
	\phi_{\rm MH}(u,\,v)&=& \frac{1}{2}\, \left\langle e^{i\,\hat{X}\,u}e^{i\,\hat{Z}\,v}+e^{i\,\hat{Z}\,v}e^{i\,\hat{X}\,u}\right\rangle  \nonumber \\ 
	&=&\sum_{x,\,z}\, \langle{\hat{G}_Q}(x,\,z;\eta)\rangle\,\,  e^{i\,(x\, u+z\,v)}
\end{eqnarray*}
lead to an explicit evaluation of $\hat{G}_Q(x,\,z;\eta)$ from the coefficient of $e^{i\,(x\, u+z\,v)}$ in the characteristic function $\phi_{\rm MH}(u,\,v)$.

We now proceed to discuss specific examples of MH-QMO for the orthogonal spin observables of qubits, qutrits and 2-qubits.  We choose  the orthogonal spin observables corresponding to  measurements along $x$ and $z$ directions.  
\subsection{Joint measurability of orthogonal qubit observables} 

Consider the orthogonal qubit observables 
\begin{eqnarray}
	\hat{X}&=&{\sigma}_x,\ \   \hat{Z}={\sigma}_z,\ \ \left[\sigma_x,\,\sigma_z\right]=-i\,\sigma_y \nonumber \\
	\sigma_x&=&\left(\begin{array}{cc} 0 & 1 \\ 1 & 0 \end{array}\right), \ \ \sigma_y=\left(\begin{array}{cc} 0 & -i \\ i & 0 \end{array}\right),\ \  \sigma_z=\left(\begin{array}{cc} 1 & 0 \\ 0 & -1 \end{array}\right).
\end{eqnarray} 
The MH characteristic function associated with the above pair of  qubit observables  is given by    
\begin{eqnarray} 
	\label{mhqq}
	\phi_{\rm MH}(u,\,v)&=&\frac{1}{2!} \left\langle\, \left(\,  e^{i\, \sigma_x\, u}\,e^{i\, \sigma_z\, v}+e^{i\, \sigma_z\, v}\,
	e^{i\, \sigma_x\, u}\right)\right\rangle \nonumber \\ 
	&=& \sum_{x,\,z}\, P_{\rm MH}(x,\,z)\,  e^{i\,(x\, u+z\,v)} \nonumber \\ 
	&=& \sum_{x,\,z}\, \langle{G_Q}(x,\,z)\rangle e^{i\,(x\, u+z\,v)}.
\end{eqnarray}
On simplification we obtain  
\begin{equation}
	\phi_{\rm MH}(u,\,v)=\langle I_2\rangle\, \cos\,u \cos\,v+i\langle \sigma_x\rangle\,\sin\,u \cos\,v+i\langle \sigma_z\rangle\,\cos\,u \sin\,v.
\end{equation}
The qubit MH-QMO $\{\hat{G}^{\rm qubit}_Q(x,z)\}$ is then identified using  the coefficient of $e^{i(xu+zv)}$ i.e.,  
\begin{equation}
	\label{noeta}
	\hat{G}^{\rm qubit}_Q(x,\,z)=\frac{1}{4}\left(I_2+\,x\,\sigma_x+\,z\sigma_z\right), \ \ x,\, z=\pm 1.
\end{equation}
\begin{itemize}
	\item[$\bullet$] We  construct a one-parameter family of generalized unsharp measurements $\{\hat{G}^{\rm qubit}_Q(x,\,z;\eta)\}$ by replacing~\cite{ft2} 
	$\hat{X}~=~\sigma_x$ by  $\eta\,\hat{X}~=~\eta\sigma_x$ and  $\hat{Z}~=~\sigma_z$ by  $\eta\,\hat{Z}~=~\eta\sigma_z$, where $0\leq \eta\leq 1$, to obtain  (see (\ref{noeta}))
	\begin{eqnarray}
		\label{mhqc}
		\hat{G}^{\rm qubit}_Q(x,\,z;\eta)&=&\frac{1}{4}\left(I_2+\eta\,x\,\sigma_x+\eta\,z\,\sigma_z\right) \nonumber \\ 
		&=&\frac{1}{4}\left(\begin{array}{cc} 
			1+\eta\, z  &   \eta\, x \\ 
			\eta\, x &  1-\eta\, z 
		\end{array}\right). 
	\end{eqnarray}
	\item[$\bullet$] Note that the elements of fuzzy POVM  associated with the individual measurements of the orthogonal qubit observables $\sigma_x$, $\sigma_z$ obey the conditions~\cite{nc,Hein2,LSW,Ravi,ARU}
	\begin{eqnarray}
		\sum_{x=\pm 1} \hat{E}^{\rm qubit}_\eta(x)=I_2,\ \ && \hat{E}^{\rm qubit}_\eta(x)\geq 0 \nonumber \\
		\sum_{z=\pm 1} \hat{F}^{\rm qubit}_\eta(z)=I_2,\ \ && \hat{F}^{\rm qubit}_\eta(z)\geq 0. 
	\end{eqnarray}   
	\item[$\bullet$] The eigenvalues of $\hat{G}^{\rm qubit}_Q(x,\,;\eta)$ are readily found to be 
	\begin{equation}
		\lambda_\pm=\frac{1}{4}(1\pm \eta\sqrt{2})
	\end{equation} 
	and it is seen that the least eigenvalue $\lambda_{-}<0$ when 
	\begin{equation}
		\label{0707}
		\eta>\frac{1}{\sqrt{2}}\approx 0.707.
	\end{equation}
	Thus $\{\hat{G}^{\rm qubit}_Q(x=\pm 1,\,z=\pm 1;\eta)\}$ is positive  in the range  $0\leq\eta\leq \frac{1}{\sqrt{2}}$. In other words  the MH-QMO $\{\hat{G}^{\rm qubit}_Q(x,\,z;\eta)\}$ with $\eta\in (0,\,\frac{1}{\sqrt{2}}\approx 0.707)$ turns out  to be a valid parent  POVM for joint measurements of the orthogonal qubit observables $\sigma_x$, $\sigma_z$. It may be seen that the {\em degree of compatibility}  $\eta_{\rm qubit}=\frac{1}{\sqrt{2}}$ matches perfectly with the results for joint measurability of a pair of orthogonal qubit  observables~\cite{LSW,Ravi,ARU,seeta_icqf}.
\end{itemize}
\subsection{Joint measurability of orthogonal qutrit observables}  
%
Consider the qutrit spin observables 
\begin{equation}
	\label{xzqutrit}
	\hat{X}=\frac{1}{\sqrt{2}}\left(\begin{array}{ccc} 0 & 1 & 0 \\ 1 & 0 & 1 \\ 0 & 1 & 0 \end{array}\right), \ \ \ \ \hat{Z}=\left(\begin{array}{ccc} 1 & 0 & 0 \\ 0 & 0 & 0 \\ 0 & 0 & -1 \end{array}\right),
\end{equation}
which are x, z components of spin-1 quantum system respectively.   

In order to evaluate the MH characteristic function  $$\phi_{\rm MH}(u,\,v)=\frac{1}{2}\, \left\langle \left\{ e^{iX\,u},\, e^{iZ\,v}\right\}\right\rangle=\frac{1}{2}\, \left\langle e^{iX\,u}e^{iZ\,v}+e^{iZ\,v}e^{iX\,u}\right\rangle$$ and to identify MH-QMO $\{\hat{G}_Q(x,\,z;\eta), x,z=\pm1, 0; 0\leq\eta\leq 1\}$ associated with the pair of qutrit spin observables $(\hat{X},\hat{Z})$ we prescribe the following steps. 
\begin{itemize}
	\item[$\bullet$] We employ the 2-qubit basis $\left\{ \vert 0_10_2\rangle,\,
	\vert 0_11_2\rangle,\,
	\vert 1_10_2\rangle,\,\vert 1_11_2\rangle \right\}$, instead of the qutrit basis~\cite{Var}  
	$\left\{ \vert jm\rangle,\ j=1,\,0,\ m=-j \ \mbox{to}\ j \right\}$ to express $\hat{X}$, $\hat{Z}$  as 
	\begin{equation}
		\hat{U}^\dagger_{\rm CG}(\hat{X}\oplus 0)\hat{U}_{\rm CG}= \frac{1}{2}(I_2\otimes \sigma_x+\sigma_x \otimes I_2), \ \
		\hat{U}^\dagger_{\rm CG}(\hat{Z}\oplus 0)\hat{U}_{\rm CG}=\frac{1}{2}(I_2\otimes \sigma_z+\sigma_z \otimes I_2)  
	\end{equation} 
	Here the symbol $\oplus$ denotes direct sum and the unitary matrix $\hat{U}_{\rm CG}$ given by~\cite{Var} 
	\begin{equation}
		\label{ucg}
		\hat{U}_{\rm CG}=\left(\begin{array}{cccc} 1 & 0 & 0 & 0 \\ 0 & \frac{1}{\sqrt{2}} & \frac{1}{\sqrt{2}} & 0 \\ 0 & 0 & 0 & 1 \\ 0 & \,\frac{1}{\sqrt{2}} & 
			-\frac{1}{\sqrt{2}} & 0
		\end{array}\right)
	\end{equation}
	is the Clebsch-Gordan coefficient matrix connecting the product angular momentum  basis 
	$\left\{ \vert j_1=\frac{1}{2} m_1\rangle\otimes \vert j_2=\frac{1}{2} m_2\rangle, \ m_1,\,m_2=\frac{1}{2},-\frac{1}{2}\right\}$ 
	with the coupled angular momentum basis  $\left\{ \vert (j_1=\frac{1}{2},j_2=\frac{1}{2}) jm\rangle,\ j=1,\,0,\ m=-j \ \mbox{to}\ j \right\}$.
	
	\item[$\bullet$] The MH characteristic function $\phi_{\rm MH}(u,\,v)= \frac{1}{2}\, \left\langle e^{iX\,u}e^{iZ\,v}+e^{iZ\,v}e^{iX\,u}\right\rangle$ assumes the form 
	{\scriptsize\begin{eqnarray} 
			\label{phiqutrit}
			\phi_{\rm MH}(u,\,v)
			&=&\frac{1}{4}\left[I_2\otimes I_2\,(1+\cos\,u+\cos\,v+\cos\, u\,\cos v)+i \,(\sigma_x\otimes I_2 +
			I_2\otimes\sigma_x)\,(\sin\,u+\sin\,u\cos\,v)\right.
			\nonumber \\
			& +& \left. i (\sigma_z\otimes I_2 +I_2\otimes\sigma_z)(\sin\,v+\sin\,v\cos\,u) 
			-\sigma_x\otimes \sigma_x\,(1-\cos\,u+\cos\,v-\cos\,u\,\cos\,v) \right.\nonumber \\ 
			&- & \left.  \sigma_y\otimes \sigma_y\, (1-\cos\,u-\cos\,v+\cos\,u\,\cos\,v) - \sigma_z\otimes \sigma_z\, (1+\cos\,u-\cos\,v-\cos\,u\,\cos\,v)\right. \nonumber \\
			&-& \left.(\sigma_x\otimes \sigma_z+\sigma_z\otimes \sigma_x)\, \sin\,u\,\sin\,v\right] \nonumber \\ 
			&=&\sum_{x,z=\pm 1, 0}\, \left\langle \hat{G}_{Q}(x,z)\right\rangle\, e^{i\,(x\,u+z\,v)}.
	\end{eqnarray}}
	\item[$\bullet$] Collecting the coefficients of  $e^{i(xu+zv)},\ x,z=\pm 1, 0$, we obtain the elements of  MH-QMO: 
	\begin{eqnarray}
		\label{mhqutrit} 
		\hat{G}_{Q}(0,\,0)&=&\frac{1}{4}\left[ I_2\otimes I_2 -\sigma_x\otimes \sigma_x-\sigma_z\otimes \sigma_z
		-\sigma_y\otimes \sigma_y \right] \nonumber \\ 
		\hat{G}_{Q}(x,\,0)&=&\frac{1}{8}\left[ I_2\otimes I_2 +\sigma_x\otimes \sigma_x+\sigma_y\otimes \sigma_y-
		\sigma_z\otimes \sigma_z+x\, (I\otimes \sigma_x+\sigma_x\otimes I)
		\right] \nonumber \\ 
		\hat{G}_{Q}(0,\,z)&=&\frac{1}{8}\left[ I_2\otimes I_2 -\sigma_x\otimes \sigma_x+\sigma_z\otimes \sigma_z
		+\sigma_y\otimes \sigma_y+z\, (I\otimes \sigma_z+\sigma_z\otimes I) \right] \nonumber \\
		\hat{G}_{Q}(x,\,z)&=&\frac{1}{16}\left[ I_2\otimes I_2 + \sigma_x\otimes \sigma_x+\sigma_z\otimes \sigma_z
		-\sigma_y\otimes \sigma_y+z\,(I\otimes \sigma_z+\sigma_z\otimes I) \right. \nonumber \\
		& &\left. +x(I\otimes \sigma_x+\sigma_x\otimes I)+ x\,z (\sigma_x\otimes \sigma_z+\sigma_z\otimes \sigma_x) \right].
	\end{eqnarray} 
	\item[$\bullet$] Then we replace $\sigma_x\longrightarrow\eta\,\sigma_x$ and  $\sigma_z\longrightarrow\eta\,\sigma_z$, $0\leq \eta \leq 1$ in (\ref{mhqutrit}) to obtain  elements of the one-parameter fuzzy MH-QMO~\cite{ft3}	
	{\scriptsize\begin{eqnarray}
			\label{mhqutriteta} 
			\hat{G}_Q(0,0;\eta)&=&\frac{1}{4}\left[ I_2\otimes I_2 -\eta^2\left(\sigma_x\otimes \sigma_x+\sigma_z\otimes \sigma_z
			+\eta^2(\sigma_y\otimes \sigma_y)\right) \right] \nonumber \\ 
			\hat{G}_Q(x=\pm 1,\,0;\eta)&=&\frac{1}{8}\left[ I_2\otimes I_2 +\eta^2\left(\sigma_x\otimes \sigma_x+\eta^2(\sigma_y\otimes \sigma_y)-
			\sigma_z\otimes \sigma_z\right)+\eta\,x(I\otimes \sigma_x+\sigma_x\otimes I)
			\right] \nonumber \\ 
			\hat{G}_Q(0,\,z=\pm 1;\eta)&=&\frac{1}{8}\left[ I_2\otimes I_2 -\eta^2\left(\sigma_x\otimes \sigma_x-(\sigma_z\otimes \sigma_z)
			-\eta^2(\sigma_y\otimes \sigma_y)\right)+\eta\,z(I\otimes \sigma_z+\sigma_z\otimes I) \right] \nonumber \\
			\hat{G}_Q(x=\pm 1,\,z=\pm 1;\eta)&=&\frac{1}{16}\left[ I_2\otimes I_2 + \eta^2\left(\sigma_x\otimes \sigma_x+\sigma_z\otimes \sigma_z
			-\eta^2(\sigma_y\otimes \sigma_y)\right)+\eta\,z(I\otimes \sigma_z+\sigma_z\otimes I) \right. \nonumber \\
			& &\left. +\eta\,x(I\otimes \sigma_x+\sigma_x\otimes I)+ \eta^2\,x\,z (\sigma_x\otimes \sigma_z+\sigma_z\otimes \sigma_x) \right].
	\end{eqnarray}} 
	\item[$\bullet$] We then transform the $4\times 4$ matrices $\hat{G}_Q$'s of (\ref{mhqutriteta})  back to the qutrit basis  using the unitary transformation (\ref{ucg}) i.e.,  
	\begin{equation}
		\hat{U}_{\rm CG}\hat{G}_Q(x,z;\eta)\hat{U}^\dagger_{\rm CG}=\left(\begin{array}{lr} \hat{G}_Q^{\rm qutrit}(x,z;\eta) & 0_{3\times 1} \\  0_{1\times 3} & \hat{G}^{\rm singlet}(x,z;\eta)\end{array}\right)  
	\end{equation}
	where 
	\begin{eqnarray}
		\label{qutritG}
		\hat{G}_Q^{\rm qutrit}(0,0;\eta)&=&\frac{1}{4} \left(\begin{array}{ccc} 1-\eta^2 & 0 &
			\eta^2(\eta^2-1) \\  0 & 1-\eta^4 & 0  \\ \eta^2(\eta^2-1) & 0 & 1-\eta^2 \end{array}\right) \nonumber \\
		\hat{G}_Q^{\rm qutrit}(x=\pm 1,0;\eta)&=&\frac{1}{8} \left(\begin{array}{ccc} 1-\eta^2 & 
			\sqrt{2}x\eta & \eta^2(1-\eta^2) \\ \sqrt{2}x\eta & (1+\eta^2)^2 & \sqrt{2}x\eta \\ \eta^2(1-\eta^2) & \sqrt{2}x\eta & 1-\eta^2 \end{array}\right) \\
		\hat{G}_Q^{\rm qutrit}(0,z=\pm 1;\eta)&=&\frac{1}{8} \left(\begin{array}{ccc} 1+2z\eta+\eta^2 & 
			0 & -\eta^2(1+\eta^2) \\ 0 & (\eta^2-1)^2  & 0 \\ -\eta^2(1+\eta^2) & 0 & 1-2z\eta+\eta^2 \end{array}\right), \nonumber \\
		\hat{G}_Q^{\rm qutrit}(x=\pm 1,z=\pm 1;\eta)&=&\frac{1}{16} \left(\begin{array}{ccc} 1+2z\eta+\eta^2 & 
			\sqrt{2}x\eta(1+z\eta) & \eta^2(1+\eta^2) \\ \sqrt{2}x\eta(1+z\eta) & 1-\eta^4 & x\sqrt{2}\eta(1-z\eta) \\ \eta^2(1+\eta^2) & \sqrt{2}x\eta(1-z\eta) & 1-2z\eta+\eta^2 \end{array}\right)\nonumber  
	\end{eqnarray}
	are the elements of qutrit MH-QMO. 
	\item[$\bullet$]  The  qutrit POVM $\{\hat{E}_\eta^{\rm qutrit}(x),x=\pm 1, 0,\}$ associated with fuzzy measurement of the observable $\hat{X}$ is then obtained using the qutrit MH-QMO (see (\ref{qutritG}) as follows:    
	\begin{eqnarray} 
		\label{ex3}
		\hat{E}_\eta^{\rm qutrit}(x=\pm 1)&=&\sum_{z=\pm 1,\,0}\,\hat{G}_Q^{\rm qutrit}(x,\,z;\eta)\nonumber \\
		&=&\frac{1}{4} \left(\begin{array}{ccc} 1 & 
			\sqrt{2}x\eta & \eta^2 \\ \sqrt{2}x\eta & 1+\eta^2 & \sqrt{2}x\eta \\ \eta^2 & \sqrt{2}x\eta & 1 \end{array}\right) \nonumber \\
		\hat{E}_\eta^{\rm qutrit}(x=0)&=& \sum_{z=\pm 1,\,0}\,G_Q^{\rm qutrit}(0,z;\eta)\nonumber \\
		&=&\frac{1}{2} \left(\begin{array}{ccc} 1 & 
			0 & -\eta^2 \\ 0 & 1-\eta^2 & 0 \\ -\eta^2 & 0 & 1 \end{array}\right)
	\end{eqnarray} 
	\item[$\bullet$] Similarly the  qutrit POVMs for the measurement of $\hat{Z}$  take the form 
	\begin{eqnarray}
		\label{fx3}
		\hat{F}_\eta^{\rm qutrit}(z=\pm 1)&=&\sum_{x=\pm 1,\,0}\,\hat{G}_Q^{\rm qutrit}(x,z;\eta) \nonumber \\ 
		&=&\frac{1}{4} \left(\begin{array}{ccc} 1+2z\eta+\eta^2 & 
			0 & 0 \\ 0 & 1-\eta^2 & 0 \\ 0 & 0 & 1-2z\eta+\eta^2 \end{array}\right) \nonumber \\
		\hat{F}_\eta^{\rm qutrit}(z=0)&=&\sum_{x=\pm 1,\,0}\,\hat{G}_Q^{\rm qutrit}(x,\,0;\eta) \nonumber \\ 
		&=&\frac{1}{2} \left(\begin{array}{ccc} 1-\eta^2 & 
			0 & 0 \\ 0 & 1+\eta^2 & 0 \\ 0 & 0 & 1-\eta^2 \end{array}\right) 
	\end{eqnarray}
	\item[$\bullet$] Note that $\{\hat{E}_\eta^{\rm qutrit}(x)\}, \ \{\hat{F}_\eta^{\rm qutrit}(z)\}$ obey the properties  
	\begin{eqnarray}
		\sum_{x=\pm 1, 0}\, \hat{E}_\eta^{\rm qutrit}(x)=I_3,\ \ \  \hat{E}_\eta^{\rm qutrit}(x)\geq 0 \nonumber \\
		\sum_{z=\pm 1, 0}\, \hat{F}_\eta^{\rm qutrit}(z)=I_3,\ \ \  \hat{F}_\eta^{\rm qutrit}(z)\geq 0 
	\end{eqnarray} 
	and they correspond to  one-parameter fuzzy measurements of the observables $\hat{X},\, \hat{Z}$ respectively. 
\end{itemize}
Now we proceed to evaluate the eigenvalues of the  joint MH-QMO \break $\left\{\hat{G}_Q^{\rm qutrit}(x,\,z;\eta); x,z=\pm 1,\,0\right\}$ and find that 
the lowest eigenvalue of both 
$\hat{G}_Q^{\rm qutrit}~(x~=~\pm~1, 0;\eta)$ and 
$\hat{G}_Q^{\rm qutrit}(0,z=\pm 1;\eta)$ is given by 
\begin{equation}
	\lambda(\eta)=\frac{1+\eta^2-\eta\sqrt{4+\eta^2(1+\eta^2)^2}}{8}
\end{equation} 
which assumes  negative values (see Fig.1) when $\sqrt{\sqrt{2}-1}<\eta\leq 1$.
\begin{figure}[h]
	\label{mhqtm}
	\begin{center}
		\includegraphics*[width=4in,keepaspectratio]{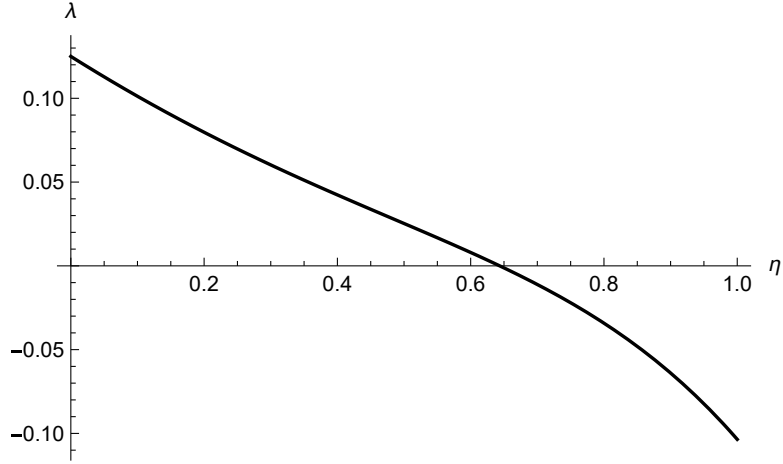}
		\caption{A plot of the lowest eigenvalue $\lambda(\eta)=\frac{1+\eta^2-\eta\sqrt{4+\eta^2(1+\eta^2)^2}}{8}$ of the elements   
			$\hat{G}_Q^{\rm qutrit}(0,z;\eta)$, for $z=\pm 1$ and $\hat{G}_Q^{\rm qutrit}(x,0;\eta)$ for $x=\pm 1$ of the  MH-QMO as a function of the unsharpness parameter  $\eta$. It is seen that $\lambda(\eta)<0$ when  
			$\eta\geq\sqrt{\sqrt{2}-1}\approx 0.64359.$} 
	\end{center}
\end{figure}   
Moreover the eigenvalues of  $\hat{G}_Q^{\rm qutrit}(x=\pm 1,z=\pm 1;\eta)$  given by 
\begin{eqnarray}
	\label{eigQ}
	\lambda_1(\eta)=\frac{1-2\eta^2-\eta^4}{16},\  & & \  \lambda_2=\frac{1+2\eta^2-\eta\sqrt{\eta^6+8}}{16} 
\end{eqnarray} 
assume negative values when $\eta\geq\sqrt{\sqrt{2}-1}$.
\begin{figure}[h]
	\label{mhqtm}
	\begin{center}
		\includegraphics*[width=4in,keepaspectratio]{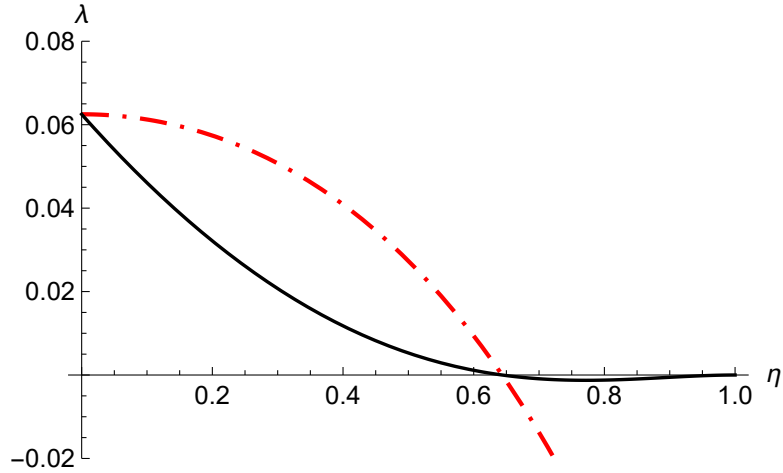}
		\caption{A plot of the eigenvalues $\lambda_1(\eta)=\frac{1-2\eta^2-\eta^4}{16}$ (dot-dashed line), $\lambda_2=\frac{1+2\eta^2-\eta\sqrt{\eta^6+8}}{16} $ 
			(undashed line) of the elements  $\hat{G}_Q^{\rm qutrit}(x=\pm 1,z=\pm 1;\eta)$  of the MH-QMO as a function of $\eta$. It is seen that  $\lambda_1(\eta),\ \lambda_2(\eta)<0$ 
			when  $\eta\geq\sqrt{\sqrt{2}-1}\approx 0.64359$.} 
	\end{center}
\end{figure} 

In other words, the qutrit MH-QMO  $\left\{\hat{G}_Q^{\rm qutrit}(x,z;\eta); x,z=\pm 1,0\right\}$ happens be a legitimate parent POVM enabling joint measurements of the orthogonal
qutrit observables $\hat{X}, \ \hat{Z}$ (see (\ref{xzqutrit})) when $0\leq \eta \leq \sqrt{\sqrt{2}-1}\approx 0.64359$. 

Note that the region of compatibility of the qutrit observables $\hat{X}$, $\hat{Z}$ i.e., 
\begin{equation}
	\label{6435}
	0\leq \eta\leq 0.64359
\end{equation} 
is smaller than  that of a qubit (see (\ref{0707})).  Thus  the incompatibility of  the pair of  spin observables $\hat{X},\ \hat{Z}$ is seen to increase with the  Hilbert space dimension. This result is consistent with the observation of Ref.~\cite{Hein2}  that  the finite dimensional position and momentum observables become more incompatible with increasing dimension. 
\subsection{Joint measurability of 2-qubit observables} 
In this section, we consider joint measurements of the set $\{\hat{X}_1,\, \hat{Z}_1; \hat{X}_2,\,\hat{Z}_2\}$ of 2-qubit observables:  
\begin{eqnarray}
	\label{2qo}
	\hat{X}_1&=&{ \sigma}_x\otimes { I}_2, \ \ \hat{Z}_1={\sigma}_z\otimes { I}_2 \nonumber \\
	\hat{X}_2&=&{I}_2\otimes { \sigma}_x, \ \ \hat{Z}_2={ I}_2\otimes { \sigma}_z.
\end{eqnarray}
The   observable pairs  $(\hat{X}_k,\hat{Z}_k), k=1,2$ represent x, z components of qubit spin observables  (expressed  in the 2-qubit computational basis $\left\{ \vert 0_10_2\rangle,\,
\vert 0_11_2\rangle,\,
\vert 1_10_2\rangle,\,\vert 1_11_2\rangle \right\}$).  

Note that 
\begin{itemize} 
	\item[$\bullet$]  the  x,z components of   qubit spin observables are shown to be jointly measurable in the  range $0\leq\eta\leq 0.707$ (see \ref{0707})) of the fuzziness parameter;  
	\item[$\bullet$] the  observable pair $(\hat{X}_1,\hat{Z}_1)$ of qubit 1 commutes with the set $(\hat{X}_2,\hat{Z}_2)$ of  qubit 2. 
\end{itemize}
So, it appears natural to anticipate that the set  $\{\hat{X}_1,\, \hat{Z}_1; \hat{X}_2,\,\hat{Z}_2\}$  of 2-qubit observables  (see \ref{2qo})) is compatible when $0\leq \eta\leq 0.707.$  With the help of an explicit construction of MH-QMO for the 2-qubit observables (\ref{2qo}) we show that even though pairwise joint measurability of the observable pairs   $(\hat{X}_1,\, \hat{Z}_1)$ and $(\hat{X}_2,\, \hat{Z}_2)$ is ensured when  $0\leq\eta \leq 0.707$, the set of all four 2-qubit observables   $\{\hat{X}_1,\, \hat{Z}_1; \hat{X}_2,\,\hat{Z}_2\}$ is  incompatible.         

The MH characteristic function $\phi_{\rm MH}(u_1,\,v_1;u_2,\,v_2)$ associated with the 2-qubit observables $\{\hat{X}_1,\,\hat{Z}_1; \hat{X}_2,\,\hat{Z}_2\}$ of (\ref{2qo})
is readily evaluated to obtain
\begin{eqnarray}
	\phi_{\rm MH}(u_1,\,v_1;u_2,\,v_2)&=& \frac{1}{2}\,\left\langle\left( e^{i\, \hat{X}_1\,u_1}e^{i\, \hat{X}_2\,u_2},\, e^{i\,\hat{Z}_1\,v_1}e^{i\,\hat{Z}_2\,v_2}\right)\right\rangle \ \   \nonumber \\
	&=& \frac{1}{2}\,\left\langle \left(e^{i(\hat{X}_1\,u_1+\hat{X}_2\,u_2)}e^{i(\hat{Z}_1\,v_1+\hat{Z}_2\,v_2)}+e^{i(\hat{Z}_1\,v_1+\hat{Z}_2\,v_2)}e^{i(\hat{X}_1\,u_1+\hat{X}_2\,u_2)}\right) 
	\right\rangle \nonumber \\
	&=&I_2\otimes I_2\,\cos u_1\cos u_2\cos v_1\cos v_2+i\,\sigma_x\otimes I_2\,\sin u_1\cos u_2\cos v_1\cos v_2  \nonumber \\
	&+&i\,I_2\otimes \sigma_x\,\cos u_1\sin u_2\cos v_1\cos v_2+i\,\sigma_z\otimes I_2\,\,\cos u_1\cos u_2\sin v_1\cos v_2 \nonumber \\
	&+&i\,I_2\otimes \sigma_{z}\,\cos u_1\cos u_2\cos v_1\sin v_2-\,\sigma_x\otimes \sigma_x\,\sin u_1\sin u_2\cos v_1\cos v_2  \nonumber \\
	&-&\sigma_z\otimes\sigma_z\,\cos u_1\cos u_2\sin v_1\sin v_2-\,\sigma_y\otimes \sigma_y\,\sin u_1\sin u_2\sin v_1\sin v_2 \nonumber\\
	&-&\,\sigma_x\otimes \sigma_z\,\sin u_1\cos u_2\cos v_1\sin v_2-\,
	\sigma_z\otimes \sigma_x\,\cos u_1\sin u_2\sin v_1\cos v_2. \nonumber \\ 
	&=& \sum_{x_1,z_1; x_2,z_2=\pm 1}\, \left\langle \hat{G}^{2-{\rm qubit}}_{Q}(x_1,z_1,x_2,z_2)\right\rangle\, e^{i\,(x_1\,u_1+z_1\,v_1+x_2\,u_2+z_2\,v_2)}
\end{eqnarray} 

We then obtain explicit form  of the elements of  2-qubit  MH-QMO    from the coefficients of  $e^{i\,(x_1\,u_1+z_1\,v_1+x_2\,u_2+z_2\,v_2)},\ \ x_1,z_1, x_2,z_2=\pm1$:
\begin{eqnarray}
	\label{gx12}
	G^{2-{\rm qubit}}_{Q}(x_1,z_1,x_2,z_2)&=&\frac{1}{16}\left(I_2\otimes I_2+x_1\,\sigma_x\otimes I_2 +z_1\,\sigma_z\otimes I_2 \right.\nonumber \\ 
	&+&  \,x_2\, I_2\otimes \sigma_x +z_2\, I_2\otimes \sigma_z\, 
	+  x_1x_2\, \sigma_x\otimes \sigma_x+z_1z_2\,\sigma_z\otimes \sigma_z\nonumber \\
	&+& \left. x_1z_2\,\sigma_x\otimes \sigma_z + x_2z_1\,\sigma_z\otimes \sigma_x\,- x_1x_2z_1z_2\, \sigma_y\otimes \sigma_y\right].
\end{eqnarray} 
In order to bring in {\em measurement fuzziness} we replace 
$\sigma_x \rightarrow \eta\sigma_x$,\  $\sigma_z \rightarrow \eta\sigma_z$
where $0\leq \eta \leq 1$. The corresponding 2-qubit fuzzy MH-QMO 
$\left\{G^{2-{\rm qubit}}_{Q}(x_1,z_1,x_2,z_2;\eta)\right\}$ is then given by
\begin{eqnarray}
	\label{gx12eta}
	G^{2-{\rm qubit}}_{Q}(x_1,z_1,x_2,z_2;\eta)&=&\frac{1}{16}\left[I_2\otimes I_2+\eta\left(x_1\, \sigma_x\otimes I_2 +z_1\, \sigma_z\otimes I_2\right) \right.\nonumber \\ 
	&+& \left. \eta \left( x_2\, I_2\otimes \sigma_x +z_2\, I_2\otimes \sigma_z\right) \right. \nonumber \\
	&+& \left. \eta^2 \left(x_1x_2\, \sigma_x\otimes \sigma_x+z_1z_2\, \sigma_z\otimes \sigma_z+x_1z_2\, \sigma_x\otimes \sigma_z \right.\right.
	\nonumber \\
	&+& \left.\left. x_2z_1\, \sigma_z\otimes \sigma_x\right)-\eta^4 x_1x_2z_1z_2\, \sigma_y\otimes \sigma_y\, \right].
\end{eqnarray} 
It is readily seen that 
\begin{eqnarray}
	(1) && \sum_{x_1,z_1,x_2,z_2=\pm 1}G^{2-{\rm qubit}}_{Q}(x_1,z_1,x_2,z_2;\eta)=I_2\otimes I_2 \nonumber \\ 
	(2) && \sum_{x_2,z_2=\pm 1}G^{2-{\rm qubit}}_{Q}(x_1,z_1,x_2,z_2;\eta)=\frac{1}{4}\left[I_2\otimes I_2+\eta\left(x_1\, \sigma_x\otimes I_2 +z_1\, \sigma_z\otimes I_2\right) \right] \nonumber \\ 
	&& \hskip 1in =G_{Q}(x_1,z_1;\eta) \nonumber \\
	&& \sum_{x_1,z_1=\pm 1}G^{2-{\rm qubit}}_{Q}(x_1,z_1,x_2,z_2;\eta)=\frac{1}{4}\left[I_2\otimes I_2+\eta\left(x_2\, I_2\otimes\sigma_x +z_2\, I_2\otimes\sigma_z\right) \right] \nonumber \\ 
	&& \hskip 1in =G_{Q}(x_2,z_2;\eta).
\end{eqnarray} 
\begin{itemize}
	\item[$\bullet$] The elements of   MH-QMOs  
	\begin{eqnarray} 
		G_{Q}(x_1,z_1;\eta)&=& \frac{1}{4}\left[I_2\otimes I_2+\eta\left(x_1\, \sigma_x\otimes I_2 +z_1\, \sigma_z\otimes I_2\right) \right],\ x_1,z_1=\pm 1  \nonumber \\ 
		G_{Q}(x_2,z_2;\eta)&=& \frac{1}{4}\left[I_2\otimes I_2+\eta\left(x_2\, I_2\otimes\sigma_x +z_2\, I_2\otimes\sigma_z\right) \right],\ x_1,z_1=\pm 1  \nonumber  
	\end{eqnarray}
	of the first and second qubit respectively turn out to be valid parent POVMs when  $0\leq \eta\leq 0.707$ ensuring that the  pairs of observables  $(\hat{X}_1,\,\hat{Z}_1)$   and $(\hat{X}_2,\,\hat{Z}_2)$ (see (\ref{2qo}))  are  pairwise measurable (in confirmity  with the results of Sec.~5.1 on joint measurability of orthogonal qubit observables).
	\item[$\bullet$] Eventhough the observable pair $(\hat{X}_1,\,\hat{Z}_1)$ of first qubit commutes with   $(\hat{X}_2,\,\hat{Z}_2)$ of the second qubit,  the total set  $(\hat{X}_1,\,\hat{Z}_1,\ \hat{X}_2,\,\hat{Z}_2)$ of 2-qubit observables (see (\ref{2qo})) is not jointly measurable in the range $0\leq \eta\leq \frac{1}{\sqrt{2}}$, which is evidenced by the non-positivity of   the associated  2-qubit MH-QMO $\{G^{2-{\rm qubit}}_{Q}(x_1,z_1,x_2,z_2;\eta)\}$.  
	\item[$\bullet$] From an explicit evaluations we find that the eigenvalues of the 2-qubit MH-QMO \break $\left\{G^{2-{\rm qubit}}_{Q}(x_1,z_1,x_2,z_2;\eta); 
	x_1,z_1,x_2,z_2=\pm 1\right\}$ (see (\ref{gx12eta})) are  same as that of the qutrit MH-QMO  $\left\{G_{Q}^{\rm qutrit}(x,z;\eta); x,z=\pm 1\right\}$ 
	(see (\ref{eigQ})). In other words the fuzzy MH-QMO 
	$\left\{G^{2-{\rm qubit}}_{Q}(x_1,z_1,x_2,z_2;\eta); x_1,z_1,x_2,z_2=\pm 1\right\}$ becomes a legitimate parent POVM of the  2-qubit observables  $\hat{X}_1,\,\hat{Z}_1,\ \hat{X}_2,\,\hat{Z}_2$  (see (\ref{2qo})) in the domain $\eta\in(0,\,\sqrt{\sqrt{2}-1}).$ So, the compatibility region of the set of 2-qubit observables $\hat{X}_1,\,\hat{Z}_1,\ \hat{X}_2,\,\hat{Z}_2$ is given by 
	\begin{equation} 
		0\leq \eta\leq \sqrt{\sqrt{2}-1}\approx 0.64359.
	\end{equation}   
\end{itemize}

\section{Summary} 
By employing the Margenau-Hill (MH) correspondence rule for associating classical functions with quantum operators, we have explicitly constructed fuzzy one parameter quasi measurement operators (QMO) to analyze the incompatibility of the non-commuting spin observables of qubits, qutrits and 2-qubit systems.  A real  parameter $0\leq \eta \leq 1$ corresponding to unsharpness of measurement  serves as an index of the minimum amount of {\em fuzziness} required so that the observables become compatible. We give a summary of our results: 
\begin{enumerate}
	\item 	MH-QMOs of a pair of orthogonal qubit observables  $\hat{X}=\sigma_x$, $\hat{X}=\sigma_z$ are positive in the range $0\leq\eta_{\rm qubit} \leq 0.707$ of the unsharpness parameter. This range of values, obtained based on the positivity of MH-QMO are in agreement with the results of  Refs.~\cite{LSW,Ravi,ARU,seeta_icqf}  for the joint measurability of a pair of orthogonal qubit measurements.
	
	\item  Based on the positivity of fuzzy MH-QMOs, which serves as a sufficient condition for joint measurability,  we recognize that a pair of qutrit  spin observables $\hat{X}$, $\hat{Z}$ (see (\ref{xzqutrit}))  are compatible when $0\leq \eta\leq 0.64359$ (see (\ref{6435})).  This region of compatibility is smaller than that for the qubit observables i.e., $0~\leq~\eta_{\rm qubit}~\leq~0.707$. Thus the pair of qutrit spin observables  $\hat{X}$, $\hat{Z}$ are  found to be  more incompatible compared to their qubit counterparts. In other words, quantum incompatibility of spin observables  appears to increase with the dimension of the Hilbert space. It  deserves future attention to recognize the region of incompatibility of spin observables in the limit $d\rightarrow \infty$. 
	
	\item  Designolle et. al.~\cite{njp19} carried out a detailed investigation to determine {\em maximal incompatibility} of a pair of observables in any finite dimensional Hilbert space. They list the incompatibility robustness value  $\eta^*= 0.6602$ and $\eta^*= 0.6830$ respectively for two specific qutrit  observable pairs (See (78), (79) of 
	Ref.~\cite{njp19}). Maximal incompatibility of sets of qudit observables has also been explored in Ref.~\cite{pra17} where numerical computation leads to  $\eta^*= 0.6794$ (see table 4 of Ref.~\cite{pra17}) for  a {\em  maximally incompatible pair} of $3$-outcome qutrit observables. The qutrit POVMs $\{\hat{E}^{\rm qutrit}_{\eta}(x), x=\pm 1, 0\}$, $\{\hat{F}^{\rm qutrit}_{\eta}(z), z=\pm 1, 0\}$ (see (\ref{ex3}), (\ref{fx3})) are incompatible when $\eta\geq 0.64359$ and they are found to outperform the ones listed in Refs.~\cite{pra17} and ~\cite{njp19}.  
	
	\item The 2-qubit observable pairs $(\hat{X}_1=\sigma_x\otimes I_2,\,\hat{Z}_1)=\sigma_z\otimes I_2)$ of qubit 1 and  $(\hat{X}_2=I_2\otimes \sigma_x$, $\hat{Z}_2=I_2\otimes \sigma_z)$ of qubit 2  are shown to be compatible in the domain  $0\leq \eta\leq \frac{1}{\sqrt{2}}$ which is in agreement with the region of compatibility of the orthogonal qubit observables $\sigma_x,\ \sigma_z$. However, eventhough the observable pair $(\hat{X}_1,\,\hat{Z}_1)$ of first qubit  commutes with the  set $(\hat{X}_2,\,\hat{Z}_2)$  of second qubit observables  the compatibility region of  the entire  set   $\{\hat{X}_1=\sigma_x\otimes I_2$, $\hat{Z}_1=\sigma_z\otimes I_2$;  $\hat{X}_2=I_2\otimes \sigma_x$, $\hat{Z}_2=I_2\otimes \sigma_z\}$ is found to be $0\leq \eta \leq 0.64359$, highlighting enhanced incompatibility of the set of qubit  observables, when they are embedded in a 2-qubit Hilbert space. 
\end{enumerate}

 We believe that our framework based on MH operator correspondence rule for joint measurability of spin observables provides  intuitive insights on quantum measurement incompatibility. Exploring incompatibility based on different operator correspondence rules~\cite{others} deserves future attention.

\section*{Acknowledgements} Sudha, ARU and IR are supported by the Department of Science and Technology(DST), India through Project No. DST/ICPS/QUST/Theme-2/2019 (Proposal Ref. No. 107).  HSK acknowledges the support of NCN through grant SHENG (2018/30/Q/ST2/00625).     


\begin{thebibliography}{0}
\bibitem{ft1}  It may be noted that once the fuzzy POVMs associated with the non-commuting observables from the Margenau-Hill quasi-measurement operators are obtained one may carry on with the computation of incompatibility robustness~\cite{njp19,Hein2} to obtain similar results.
\bibitem{ft2}  Replacing the Hermitian observable $\hat{X}$ by $\eta\,\hat{X}$ indicates that the moments of the observable are {\em fuzzy} i.e.,  $\left\langle\hat{X}^k\right\rangle_{\rm fuzzy}=\eta^k\,\left\langle\hat{X}^k\right\rangle_{\rm sharp}$. It also signifies that a POVM,  constructed by mixing the sharp PV measurements of the observable $\hat{X}$ with {\em noise} characterized by the parameter $\eta$ leads to unsharp measurements of the observable $\hat{X}$.
\bibitem{ft3} Replacing $\sigma_i\longrightarrow\eta\,\sigma_i,\, i=x, z$  is consistent with changing the qutrit spin observables $\hat{X}, \ \hat{Z}$ given by (\ref{xzqutrit})) with $\eta\, \hat{X},\ \eta\, \hat{Z}$ respectively.  To see this, note that $\hat{X}\rightarrow \eta\, \hat{X}, \ \hat{Z}\rightarrow\eta\, \hat{Z}$ is equivalent to replacing the corresponding spin observables  $\frac{1}{2}\left(\sigma_i\otimes I_2+I_2\otimes\sigma_i\right), i=x,z$ (i.e., when the spin observables are expressed in the two-qubit basis)  by  $\frac{\eta}{2}\left(\sigma_i\otimes I_2+I_2\otimes\sigma_i\right), i=x,z$, which in turn corresponds to changing  $\sigma_x\longrightarrow\eta\,\sigma_x,\  \sigma_z\longrightarrow\eta\,\sigma_z$.
\end{thebibliography}
\end{document}